\documentclass[epj,twocolumn]{webofc}
\usepackage[varg]{txfonts}   
\woctitle{}
\begin{document}
\title{Probing non-standard top-quark couplings 
via optimal-observable analyses at LHC
}
%
%
\author{Zenr\=o Hioki\inst{1}\fnsep\thanks{\email{hioki@ias.tokushima-u.ac.jp}}
        \and
        Kazumasa Ohkuma\inst{2}\fnsep\thanks{\email{ohkuma@fukui-ut.ac.jp}}
}

\institute{Institute of Theoretical Physics, University of Tokushima,
Tokushima 770-8502, Japan
\and
Department of Information Science, Fukui University of Technology,
Fukui 910-8505, Japan
          }

\abstract{%
Focusing attention on top-quark pair production and its decay processes
at the LHC, non-standard top-quark couplings are studied based on the
effective Lagrangian constructed with $SU(3)\times SU(2)\times U(1)$ 
invariant dimension-6 operators. The optimal-observable analysis is
carried out for the charged-lepton distributions in $pp \to t \bar{t} X
\to \ell^+ X'$ ($\ell=e$ or $\mu$) in order to estimate the expected
statistical uncertainties in measurements of those non-standard
$t\bar{t}g$ and $tbW$ couplings that contribute to this process in the
leading order. 

}
\maketitle
%
\section{Introduction}
\label{intro}
Almost twenty years  have passed since the top quark was discovered at the
Tevatron \cite{topdisc,Abachi:1995iq}. Even today, however, this quark is
still the heaviest in the experimentally-confirmed elementary particles.
Therefore, precise measurements of the top quark are expected to be one of 
the most promising ways for finding possible signals from new physics beyond 
the standard model in the current situation where there has been no 
discovery of non-standard particles~\cite{Kamenik:2011wt}.
The Large Hadron Collider (LHC) is now the unique facility 
for those top-quark studies in stead of the Tevatron.

In any processes, non-standard signals mean non-negligible differences 
between actual experimental data and the corresponding standard-model
prediction. Such deviations might be originated from quantum effects of
new particles which are not belonging to the standard-model framework.
Those quantum effects are usually characterized as new form factors
corresponding to the non-standard couplings. The size of the form factors
are calculable if you take a specific model which is an extension of the
standard model. We are also able to treat those form factors as parameters
in the manner of effective-Lagrangian
approach, 
if we would like to perform analyses as model-independently as possible.
The latter is the strategy which we have been taking for our new-physics
search in top-couplings~\cite{Hioki:2009hm,HIOKI:2011xx,Hioki:2012vn}.

In this note, we focus on our latest work~\cite{Hioki:2012vn} and show
the studies on the top-quark production and its semileptonic decay process
$pp\to t\bar{t} X \to \ell^+ X'$ ($\ell=e$ or $\mu$) at the LHC aiming
to estimate the statistical significance of measurable non-standard
top-quark couplings. Here, non-standard interactions which are originated
from the quantum effects of non-standard particles are described by
$SU(3)\times SU(2)\times U(1)$ invariant dimension-6 operators, and
parameterized as non-standard top couplings. The statistical significances
are estimated through the optimal-observable analysis.
%
\section{Strategy}
\label{sec-1}
In this section, we explain briefly the effective Lagrangian and 
the optimal-observable analysis, both of which play important roles in our
analysis.
\subsection{Effective Lagrangian}
The effective-Lagrangian approach, in which its low-energy form reproduces
the standard model, is one of the most general methods to describe new-physics
phenomena when the energy of our experimental facility is not high enough
to produce new particles. Assuming any non-standard-model particles too heavy
to appear as real ones, the effective Lagrangian is given as
\begin{equation}\label{eq:efflag}
{\cal L}_{\rm eff}={\cal L}_{\rm SM} 
+ \frac{1}{{\mit\Lambda}^2} \sum_i
\left(\,  {C_i \cal O}_i + C_i^* {\cal O}_i^{\dagger} \,\right),
\end{equation}
where ${\cal L}_{\rm SM}$ is the standard-model Lagrangian, ${\cal O}_i$ mean
$SU(3)\times SU(2)\times U(1)$ gauge-invariant operators of mass-dimension 6
involving only the standard-model fields~\cite{Buchmuller:1985jz, Arzt:1994gp}
and their coefficients $C_i$ parameterize quantum effects of new particles
at an energy less than the assumed new-physics scale 
${\mit\Lambda}$.\footnote{Since dimension-5 operators violate the lepton-number   
    conservation, they are not treated hereafter. Therefore, we deal with
    dimension-6 operators, which give the largest contributions in relevant
    processes.}\
In this framework, all the form factors related to $C_i$ are dealt with as
constant parameters, without supposing any specific new-physics models.

A full list of the dimension-6 operators were initially presented by Buchm\"{u}ller
and Wyler~\cite{Buchmuller:1985jz}, and also by Arzt {\it et al.} \cite{Arzt:1994gp}.
However, it was pointed out that some operators there are related with others
through equations of motion, which means that they are not independent of
each other~\cite{Grzadkowski:2003tf}. Then, the whole related operators were
rearranged to get rid of this redundancy
in Refs.~\cite{AguilarSaavedra:2008zc,Grzadkowski:2010es}.

Following the notation of Ref.~\cite{AguilarSaavedra:2008zc},
the relevant effective Lagrangian which describes non-standard
interactions of the third generation for the parton-level process
$q \bar{q} /gg  \to t \bar{t} \to \ell^+ X$ is given in~\cite{HIOKI:2011xx} as
\begin{alignat}{1}\label{eq:efflag_3rd}
 {\cal L}_{\rm eff} &={\cal L}_{t\bar{t}g,gg}+{\cal L}_{tbW}:  \\
 &  {\cal L}_{t\bar{t}g,gg} = -\frac{1}{2} g_s
    \sum_a \Bigl[\,\bar{\psi}_t(x)\lambda^a \gamma^{\mu}
    \psi_t(x) G_\mu^a(x)
    \bigl. \nonumber\\
  &\phantom{====}-\bar{\psi}_t(x)\lambda^a\frac{\sigma^{\mu\nu}}{m_t}
    \bigl(d_V+id_A\gamma_5\bigr)
  \psi_t(x)G_{\mu\nu}^a(x)\,\Bigr], \\
 &  {\cal L}_{tbW}  = -\frac{1}{\sqrt{2}}g 
  \Bigl[\,\bar{\psi}_b(x)\gamma^\mu(f_1^L P_L + f_1^R P_R)\psi_t(x)
   W^-_\mu(x) \Bigr. \nonumber\\
  &\phantom{====}+\bar{\psi}_b(x)\frac{\sigma^{\mu\nu}}{M_W}(f_2^L P_L + f_2^R P_R)
   \psi_t(x)\partial_\mu W^-_\nu(x) \,\Bigr], 
\end{alignat}
where $g_s$ and $g$ are the $SU(3)$ and $SU(2)$ coupling constants, 
$P_{L/R}\equiv(1\mp\gamma_5)/2$,
$d_V, d_A$ and $f_{1,2}^{L,R}$ are form factors defined as
\begin{alignat}{2}\label{eq:dvdadef}
 d_V &\equiv \frac{\sqrt{2}v m_t}{g_s {\mit\Lambda}^2} {\rm Re}(C^{33}_{uG\phi}),
 & \quad  d_A&\equiv
    \frac{\sqrt{2}v m_t}{g_s {\mit\Lambda}^2} {\rm Im}(C^{33}_{uG\phi}), \nonumber\\
  f_1^L&\equiv V_{tb}+C^{(3,33)*}_{\phi q}\frac{v^2}{{\mit\Lambda}^2},  
  & \quad  f_1^R&\equiv C^{33*}_{\phi \phi}\frac{v^2}{2{\mit\Lambda}^2},  \\
  f_2^L&\equiv -\sqrt{2} C^{33*}_{dW}\frac{v^2}{{\mit \Lambda}^2},
  & \quad  f_2^R&\equiv -\sqrt{2} C^{33}_{uW}\frac{v^2}{{\mit\Lambda}^2}, \nonumber
\end{alignat}
with $v$ being the Higgs vacuum expectation value
and $V_{tb}$ being ($tb$) element of Kobayashi--Maskawa matrix.
 In particular, $d_V$ and $d_A$ are the
so-called chromo\-magnetic- and chromo\-electric-dipole moments, respectively.

In the following calculations, we use the above effective Lagrangian for
top-quark interactions and the usual standard-model Lagrangian 
for the other interactions which are not affected by top quarks hereafter.

\subsection{Optimal-observable analysis}
The optimal-observable analysis is a method that could systematically estimate
the expected statistical uncertainties of measurable parameters
\cite{Atwood:1991ka, Davier:1992nw, Diehl:1993br,Gunion:1996vv}. 
Here we give a brief review of this procedure.

First, assume that we have a cross section
\begin{equation}
\frac{d\sigma}{d\phi}(\equiv{\mit\Sigma}(\phi))=\sum_i c_i f_i(\phi),
\label{distribution}
\end{equation}
where $f_i(\phi)$ are known functions of the final-state
variables $\phi$ and $c_i$'s are model-dependent coefficients.
The goal is to determine the $c_i$'s. This can be done by using
appropriate weighting functions $w_i(\phi)$ such that $\int w_i(\phi)
{\mit\Sigma}(\phi)d\phi=c_i$. Then, we determine $w_i(\phi)$ as the
next step. In general, different choices for
$w_i(\phi)$ are possible, but there is a unique choice for which the
resultant statistical error is minimized. Such functions are given by
\begin{equation}
  w_i(\phi)=\sum_j X_{ij}f_j(\phi)/{\mit\Sigma}(\phi)\,, \label{X_def}
\end{equation}
where $X_{ij}$ is the inverse matrix of ${\cal M}_{ij}$ which
is defined as
\begin{equation}
{\cal M}_{ij}
  \equiv \int d\phi\,f_i(\phi)f_j(\phi)/{\mit\Sigma}(\phi)\,.
\label{M_def}
\end{equation}
Finally, using Eqs.~(\ref{X_def}) and (\ref{M_def}), the statistical uncertainty
of $c_i$ is obtained as
\begin{equation}
  |\delta c_i|=\sqrt{X_{ii}\,\sigma_T/N}\,, \label{delc_i}
\end{equation}
where $\sigma_T\equiv\int (d\sigma/d\phi) d\phi$ and $N$ is the total
number of events.

Let us apply this technique to the process $pp\to t\bar{t} X
\to \ell^+ X'$, expressing its differential cross section (the angular and
energy distribution of the charged-lepton $\ell^+$) which we derived in
\cite{HIOKI:2011xx} as follows:
\begin{alignat}{1}
\frac{d^2\sigma_{\ell}}{d E_{\ell} d \cos\theta_\ell} 
  &=  f_{\rm SM}(E_{\ell}, \cos\theta_\ell) 
  + d_V f_{d_V}(E_{\ell},\cos\theta_\ell) \nonumber\\
  &\phantom{\equiv ~}+ d_R f_{d_R}(E_{\ell},\cos\theta_\ell)
  + d_V^2 f_{d_V^2}(E_{\ell}, \cos\theta_\ell) \nonumber\\
  &\phantom{\equiv ~}+ d_A^2 f_{d_A^2}(E_{\ell}, \cos\theta_\ell)
  + \cdots,
\end{alignat}
where $f_{\rm SM}(E_{\ell}, \cos\theta_\ell)$ denotes the standard-model contribution,
all the other $f_I(E_{\ell}, \cos\theta_\ell)$ describe the non-standard model terms
corresponding to their coefficients, and $d_R$ is defined as 
\begin{equation}\label{eq:drdef}
  d_R\equiv \frac{M_W}{m_t}{\rm Re}(f_2^R) .
\end{equation}
The explicit forms of $f_I(E_{\ell}, \cos\theta_\ell)$ at the parton level are
easily found in the relevant formulas in~\cite{HIOKI:2011xx}.

Since the magnitude of $d_V$ and $d_A$ has been shown
small~\cite{HIOKI:2011xx,Hioki:2012vn},
we neglect any contribution from terms quadratic (or higher)
in those non-standard model parameters
hereafter.\footnote{The quadratic terms of $f_{1,2}^{L,R}$ have already
    been neglected in our previous papers. Indeed,
    studies at the Tevatron suggest that those contributions are tiny:
    See~\cite{Aaltonen:2012rz} for the latest data.}
Note that all $d_A$ contributions disappear under this linear approximation,
because there is no term proportional to $d_A$.
Thus, the angular and energy distributions of a decay lepton are written as
\begin{alignat}{1}
 \frac{d\sigma_{\ell}}{d \cos\theta_{\ell}} 
  &=  g_1(\cos\theta_\ell) + d_V ~g_2(\cos\theta_\ell),\\
 \frac{d\sigma_{\ell}}{d E_\ell} 
  &=  ~h_1(E_{\ell}) + d_V ~h_2(E_{\ell}) 
  + d_R~h_3(E_{\ell}),
\end{alignat} 
where $g_i(\cos\theta_\ell)$ and $h_i(E_{\ell})$ are given by
\begin{alignat*}{2}
  g_i(\cos\theta_\ell)&= \int d E_\ell~f_I(E_\ell,\cos\theta_\ell),  & \quad  \\
  h_i(E_{\ell})&=\int d \cos\theta_\ell~f_I(E_\ell,\cos\theta_\ell)
\end{alignat*}
with $i=1,2$ and 3 corresponding to $I={\rm SM}$, $d_V$ and $d_R$, respectively.
Here should be one comment about the angular distribution:
We can thereby probe exclusively the $d_V$ term, since any contribution from
$d_R$ disappears within our approximation as a result of the decoupling theorem found
in~\cite{Grzadkowski:1999iq,Rindani:2000jg,Godbole:2006tq}.

We are now ready to calculate the following matrices:
\begin{alignat}{1}
 M_{ij}^c &\equiv \int d\cos\theta_{\ell} 
\frac{g_i(\cos\theta_\ell) g_j(\cos\theta_\ell)}
     {g_1(\cos\theta_\ell)}\ \ \ \ (i,\,j=1,\,2), \\
  M_{ij}^E&\equiv \int d E_\ell
  \frac{h_i(E_{\ell}) h_j(E_{\ell})}
     {h_1(E_{\ell})}\ \ \ \ (i,\,j=1,\,2,\,3)
\end{alignat}
and their inverse matrices $X_{ij}^{c,E}$, all of which are apparently symmetric.
Then the statistical uncertainties for the measurements of couplings
$d_V$ and $d_R$ could be estimated by
\begin{equation}\label{eq:uncertainty1}
| \delta d_V |
  = \sqrt{X_{22}^{c}\sigma_{\ell}/N_{\ell}}=\sqrt{X_{22}^{c}/L}
\end{equation}
through the angular distribution, and
\begin{eqnarray}
  &&| \delta d_V |= \sqrt{X_{22}^{E}\sigma_{\ell}/N_{\ell}}
  =\sqrt{X_{22}^{E}/L}\,, \label{eq:uncertainty2}\\
%
  &&| \delta d_R |= \sqrt{X_{33}^{E}\sigma_{\ell}/N_{\ell}}
  =\sqrt{X_{33}^{E}/L}\label{eq:uncertainty3}
\end{eqnarray}
via the energy distribution,
where $\sigma_{\ell}$, $N_{\ell}$ and $L$ denote
the total cross section, the number of events and the integrated luminosity
for the process, respectively.
\section{Numerical results and discussion}
Below we show the elements of $M^{c,E}$ computed for $\sqrt{s}=7,8,10$,
and 14 TeV, assuming $m_t=173$ GeV \cite{Brandt:2012ui}.
\begin{description}
 \item[(1)] The angular distribution 
 \begin{description} 
 \item[(1-1)] $\sqrt{s}=7$ TeV 
 \begin{alignat*}{3}
  M^c_{11}&= 23.102, & \quad  M^c_{12}&= -245.412,  
  & \quad  M^c_{22}&=2607.340.
\end{alignat*}
 \item[(1-2)] $\sqrt{s}=8$ TeV
  \begin{alignat*}{3}
  M^c_{11}&= 33.234, & \quad  M^c_{12}&= -353.598,  
  & \quad  M^c_{22}&= 3762.753.
\end{alignat*}
 \item[(1-3)] $\sqrt{s}=10$ TeV
\begin{alignat*}{3}
  M^c_{11}&= 59.333, & \quad  M^c_{12}&=-632.179,  
  & \quad  M^c_{22}&= 6736.735.
\end{alignat*}
 \item[(1-4)] $\sqrt{s}=14$ TeV
\begin{alignat*}{3}
  M^c_{11}&= 134.052, & \quad  M^c_{12}&= -1428.300,  
  & \quad  M^c_{22}&=15220.286.
\end{alignat*}
\end{description}
 \item[(2)] The energy distribution
 \begin{description}
 \item[(2-1)] $\sqrt{s}=7$ TeV
 \begin{alignat*}{3}
  M^E_{11}&= 23.102, & \quad  M^E_{12}&= -245.412,  
  & \quad  M^E_{13}&= 0.000,\\
  M^E_{22}&= 2607.658, & \quad  M^E_{23}&= -1.974,  
  & \quad  M^E_{33}&=15.323. 
\end{alignat*}
 \item[(2-2)] $\sqrt{s}=8$ TeV
  \begin{alignat*}{3}
  M^E_{11}&= 33.234, & \quad  M^E_{12}&= -353.598,  
  & \quad  M^E_{13}&= 0.000,\\
  M^E_{22}&= 3763.252, & \quad  M^E_{23}&= -2.880,  
  & \quad  M^E_{33}&=21.124. 
\end{alignat*}
 \item[(2-3)] $\sqrt{s}=10$ TeV
\begin{alignat*}{3}
  M^E_{11}&=59.333, & \quad  M^E_{12}&= -632.179,  
  & \quad  M^E_{13}&= 0.000,\\
  M^E_{22}&=6737.696, & \quad  M^E_{23}&= -5.120,  
  & \quad  M^E_{33}&=35.233. 
\end{alignat*}
 \item[(2-4)] $\sqrt{s}=14$ TeV
\begin{alignat*}{3}
  M^E_{11}&= 134.052, & \quad  M^E_{12}&= -1428.300,  
  & \quad  M^E_{13}&= 0.000,\\
  M^E_{22}&= 15222.443, & \quad  M^E_{23}&= -10.953,  
  & \quad  M^E_{33}&= 72.264. 
\end{alignat*}
\end{description}
\end{description}
Here, all $M_{13}^E$ became zero for the same reason as vanishing $d_R$ terms
in the angular distribution, i.e., 
the decoupling theorem~\cite{Grzadkowski:1999iq,Rindani:2000jg,Godbole:2006tq}.
Using the inverse matrices calculated from the above elements 
and Eq.(\ref{eq:uncertainty1}) -- Eq.(\ref{eq:uncertainty3}),
we can estimate the statistical uncertainties of the relevant couplings.

Before giving the results, however, we should comment about instabilities of
inverse-matrix computations: We noticed that the results fluctuate to a certain
extent (beyond our expectation) depending on to which decimal places of
$M^{c,E}$ we take into account as our input data.
Therefore we calculate $X^{c,E}$ for the above $M^{c,E}$ and also for what
are obtained by rounding those $M^{c,E}$ off to two decimal places.
We then use their mean values to get
$\delta d_V$ and $\delta d_R$ with ``errors'', which are the differences
between the mean values and the maximum/minimum.

Let us explain these treatments more specifically by taking (1-1) as an example:
For those different inputs, we get
\begin{eqnarray*}
&&\sqrt{X_{22}^c} = 1.73\ \ {\rm for}\ \ M_{11}^c=23.102,\,\ \cdots,      \\
&&\phantom{\sqrt{X_{22}^c}}
                  = 2.57\ \ {\rm for}\ \ M_{11}^c=23.10,\phantom{2}\ \cdots, 
\end{eqnarray*}
which lead to the mean value 2.15, the maximum 2.57, and the minimum 1.73, and
resultant errors $\pm 0.42$, which are derived by $2.57-2.15$ and $1.73-2.15$.

We also have to explain how we can take into account QCD higher-order corrections:
All the numerical computations of $M_{ij}^{c,E}$ in this section were done with the
tree-level formulas. In order to include QCD corrections there, we multiply the
tree cross sections by the $K$-factor. This factor disappears in the combination
$X_{ii}^{c,E}\sigma_{\ell}$ and remains only in $N_{\ell}\,(=L\sigma_{\ell})$ when
we estimate $\delta d_{V,R}$. Therefore the luminosity $L$ discussed in the following
should be understood as an effective one including $K$ (and also the lepton detection
efficiency $\epsilon_\ell$).

We now show the whole results in Table~\ref{tab:ang} and Table \ref{tab:ene}
for the angular and energy distributions, respectively. 
The uncertainties we encountered here are hard to eliminate, but still the results
will tell us the necessary luminosity for reaching the precision
which we aim to realize.
For example, we need at least $L \simeq 1500~{\rm pb}^{-1}(=1.5 ~{\rm fb}^{-1})$
in order to achieve $|\delta d_V|\simeq O(10^{-2})$ in case of measuring
the angular distribution at the LHC whose colliding energy is 8 TeV.
As mentioned, those $L$ should be divided by $K \simeq 1.5$ and $\epsilon_\ell$.
If we assume $\epsilon_\ell = 0.5$, the resultant $L$ increases slightly, which
however hardly affects our conclusion.
%
%
%
\vspace*{0.5cm}
\begin{table}[h]
\begin{center}
\caption{Estimated statistical uncertainties of $d_V$ from the angular distribution of 
a decay lepton.}\label{tab:ang}
\begin{tabular}{ cl }
\hline\vspace*{-0.35cm}\\
$\sqrt{s}$~[TeV] & $|\delta d_V|\times \sqrt{L}$ \\ \hline
 7&$2.15 \pm 0.42$\\ 
 8&$2.25 \pm 0.96$\\ 
 10&$1.12 \pm 0.12 $\\ 
 14&$0.73 \pm 0.02$\\ \hline
\end{tabular}

\end{center}
%
\begin{center}
\caption{Estimated statistical uncertainties of $d_V$ and $d_R$ from the energy
 distribution of a decay lepton.}\label{tab:ene}
\begin{tabular}{cll} 
\hline\vspace*{-0.35cm}\\
$\sqrt{s}$~[TeV]&$|\delta d_V|\times \sqrt{L}$ &$|\delta d_R|\times \sqrt{L}$ \\ \hline
7 &$1.86\pm 0.28$ &$0.35 \pm 0.02$ \\ 
8 &$1.70\pm 0.51$ &$0.32 \pm 0.05$ \\ 
10&$0.98\pm 0.08$ &$0.22 \pm0.01$ \\
14&$0.65 \pm 0.02$ &$0.15$ \\\hline
\end{tabular}
\end{center}
\end{table}
%
\section{Summary}
%
Using a scenario of Buchm\"{u}ller and Wyler~\cite{Buchmuller:1985jz} and 
the optimal-observable analysis,
we studied the statistical significance of possible non-standard
top-gluon couplings corresponding to $d_{V,A}$ and top-$W$ coupling
corresponding to 
$d_R$ as model-independently as possible for the current and future LHC experiments.

Since it has been known that those non-standard couplings are not so large
\cite{HIOKI:2011xx,Hioki:2012vn},\footnote{It is expected  that results of LHC
    experiments with $\sqrt{s}$=8 TeV will give stronger constraints for
    $d_{V,A}$~\cite{Hiokinp}.}\ 
we adopted a linear approximation for performing the optimal-observable analysis.
Although the $d_A$ terms disappeared under this approximation, 
we got the following knowledge about the determination accuracy of $d_V$ and $d_R$:

\begin{itemize}
 \item Measuring the energy distribution could study both $d_V$ and $d_R$ at the 
same time with a higher precision than the case of the angular distribution.
\item Since the angular distribution is affected by only $d_V$ contributions,
this one is suitable for exploring the mechanism of 
top-pair productions exclusively though the precision of $d_V$ is slightly 
lower than that from the energy distribution.
\end{itemize}

Furthermore, it was pointed out that 
there were some ambiguities in inverse-matrix computations depending on
where to round the input data (i.e., $M_{ij}^{c,E}$-elements) off, and consequently 
estimated statistical errors were affected thereby to a certain extent.
However, our conclusion about the necessary luminosity and new-physics scale
${\mit\Lambda}$ do not receive that serious influence, 
because those ambiguities do not change the order of the results.

Finally, if the LHC is going to be steadily upgraded and
$L=500 ~{\rm fb}^{-1}$ is achieved at $\sqrt{s}=14$ TeV,
$d_V$ and $d_R$ could be determined with
$|\delta d_V| \sim O(10^{-3})$  and $|\delta d_R| \sim O(10^{-4})$ respectively
from the energy distribution:
Here, 
$|\delta d_V| \sim 0.001$ $(|\delta d_R| \sim 0.0001)$ means that the
contribution from effective operator ${\cal O}_{uG\phi}^{33}~ ({\cal O}_{uW}^{33})$
is suppressed by ${\mit\Lambda} \gtrsim 7 ~(20)$ TeV 
if the center value of measured $d_V$ ($d_R$) is close to zero 
and $C^{33}_{uG\phi} ~(C_{uW}^{33}) \sim 1$, which we estimated from 
Eqs.(\ref{eq:dvdadef}) and (\ref{eq:drdef}).

On the other hand, 
concerning $d_A$ terms, which were neglected in this work, they induce CP-violating
interactions, therefore those effects would be better off being probed via 
some kind of asymmetric observables~\cite{Valencia:2013yr}.
%
%
\section*{Acknowledgments}
%
This work was partly supported by the Grant-in-Aid for Scientific Research 
No. 22540284 from the Japan Society for the Promotion of Science.
Part of the algebraic and numerical calculations were carried 
out on the computer system at Yukawa Institute for Theoretical
Physics (YITP), Kyoto University. 


\begin{thebibliography}{}
%
%
\bibitem{topdisc}
 F.~Abe {\it et al.}  (CDF Collaboration),
 Phys.\ Rev.\ Lett.\  {\bf 74} (1995) 2626 (hep-ex/9503002).
\bibitem{Abachi:1995iq}
  S.~Abachi {\it et al.}  (D0 Collaboration),
  Phys.\ Rev.\ Lett.\  {\bf 74} (1995) 2632 (hep-ex/9503003).
%
\bibitem{Kamenik:2011wt}
  For recent review
  J.F.~Kamenik, J.~Shu and J.~Zupan,
  arXiv:1107.5257 [hep-ph].
%
\bibitem{Hioki:2009hm}
  Z.~Hioki and K.~Ohkuma,
  Eur.\ Phys.\ J.\  C {\bf 65} (2010) 127 (arXiv:0910.3049 [hep-ph]);
  Eur.\ Phys.\ J.\  C {\bf 71} (2011) 1535 (arXiv:1011.2655 [hep-ph]).
%
\bibitem{HIOKI:2011xx}
  Z.~Hioki and K.~Ohkuma,
  Phys.\ Rev.\ D {\bf 83} (2011) 114045
  (arXiv:1104.1221 [hep-ph]).
%
%
\bibitem{Hioki:2012vn}
  Z.~Hioki and K.~Ohkuma,
  Phys.\ Lett.\ B {\bf 716} (2012) 310
  (arXiv:1206.2413 [hep-ph]).
\bibitem{Buchmuller:1985jz}
  W.~Buchmuller and D.~Wyler,
  Nucl.\ Phys.\  B {\bf 268} (1986) 621.
\bibitem{Arzt:1994gp}
  C.~Arzt, M.B.~Einhorn and J.~Wudka,
  Nucl.\ Phys.\  B {\bf 433} (1995) 41
  (hep-ph/9405214).
%
\bibitem{Grzadkowski:2003tf}
  B.~Grzadkowski, Z.~Hioki, K.~Ohkuma and J.~Wudka,
  Nucl.\ Phys.\  B {\bf 689} (2004) 108 (hep-ph/0310159).
%
\bibitem{AguilarSaavedra:2008zc}
  J.A.~Aguilar-Saavedra,
  Nucl.\ Phys.\  B {\bf 812} (2009) 181 (arXiv:0811.3842 [hep-ph]);
  Nucl.\ Phys.\  B {\bf 821} (2009) 215 (arXiv:0904.2387 [hep-ph]).
\bibitem{Grzadkowski:2010es}
  B.~Grzadkowski, M.~Iskrzynski, M.~Misiak and J.~Rosiek,
  JHEP {\bf 1010} (2010) 085 (arXiv:1008.4884 [hep-ph]).
%
%
%
%
\bibitem{Atwood:1991ka}
  D.~Atwood and A.~Soni,
  Phys.\ Rev.\ D {\bf 45} (1992) 2405.
\bibitem{Davier:1992nw}
  M.~Davier, L.~Duflot, F.~Le Diberder and A.~Rouge,
  Phys.\ Lett.\ B {\bf 306} (1993) 411.
\bibitem{Diehl:1993br}
  M.~Diehl and O.~Nachtmann,
  Z.\ Phys.\ C {\bf 62} (1994) 397.
\bibitem{Gunion:1996vv}
  J.F.~Gunion, B.~Grzadkowski and X.-G.~He,
  Phys.\ Rev.\ Lett.\  {\bf 77} (1996) 5172
  (hep-ph/9605326).
  %
\bibitem{Aaltonen:2012rz}
  T.~Aaltonen {\it et al.}  [CDF and D0 Collaborations],
  Phys.\ Rev.\ D {\bf 85} (2012) 071106
  (arXiv:1202.5272 [hep-ex]).
%
\bibitem{Grzadkowski:1999iq}
  B.~Grzadkowski and Z.~Hioki,
  Phys.\ Lett.\  B {\bf 476} (2000) 87 (hep-ph/9911505);
  Phys.\ Lett.\  B {\bf 529} (2002) 82 (hep-ph/0112361);
  Phys.\ Lett.\  B {\bf 557} (2003) 55 (hep-ph/0208079).
\bibitem{Rindani:2000jg}
  S.D.~Rindani,
  Pramana {\bf 54} (2000) 791 (hep-ph/0002006).
\bibitem{Godbole:2006tq}
  R.M.~Godbole, S.D.~Rindani and R.K.~Singh,
  JHEP {\bf 0612} (2006) 021 (hep-ph/0605100).
\bibitem{Brandt:2012ui}
  O.~Brandt [CDF and D0 Collaborations],
  arXiv:1204.0919 [hep-ex].
\bibitem{Hiokinp}
Z. Hioki and K. Ohkuma, in preparation.
\bibitem{Valencia:2013yr}
  G.~Valencia,
  PoS HQL {\bf 2012} (2012) 050
  (arXiv:1301.0962 [hep-ph]).
\end{thebibliography}
\end{document}